\documentclass[a4paper,amsmath,amssymb]{jpconf}
\usepackage{graphicx}
\usepackage{amsmath, amsthm, amssymb}
\begin{document}
\title{Phase Transition of Generalized Ferromagnetic Potts Model -- Effect of Invisible States --}

\author{Shu Tanaka}

\address{
Research Center for Quantum Computing, 
Interdisciplinary Graduate School of Science and Engineering, 
Kinki University.
3-4-1, Kowakae, Higashi-Osaka, Osaka 577-8502, Japan
}

\ead{shu-t@alice.math.kindai.ac.jp}

\author{Ryo Tamura}

\address{
Institute for Solid State Physics, University of Tokyo.
5-1-5, Kashiwanoha Kashiwa, Chiba 277-8581, Japan
}
\ead{r.tamura@issp.u-tokyo.ac.jp}

\author{Naoki Kawashima}
\address{
Institute for Solid State Physics, University of Tokyo.
5-1-5, Kashiwanoha Kashiwa, Chiba 277-8581, Japan
}
\ead{kawashima@issp.u-tokyo.ac.jp}

%%%%%%%%%%%%%%%%%%%%%%%%%%%%%%%%%%%%%%%%%%%%%%%%%%%%%%%
%%                                                                               Abstract                                                                                             %%
%%%%%%%%%%%%%%%%%%%%%%%%%%%%%%%%%%%%%%%%%%%%%%%%%%%%%%%
\begin{abstract}
We investigate the nature of the phase transition of the ferromagnetic Potts model with invisible states.
The ferromagnetic Potts model with invisible states can be regarded as straightforward extension of the standard ferromagnetic Potts model.
The invisible states contribute the entropy, however they do not affect the internal energy.
They also do not change the symmetry which breaks at the transition temperature.
The invisible states stimulate a first-order phase transition.
We confirm that the first-order phase transition with spontaneous $q$-fold symmetry breaking for $q=2,3$, and $4$ takes place even on two-dimensional lattice by Monte Carlo simulation.
We also find that the transition temperature decreases and the latent heat increases as the number of invisible states increases.
\end{abstract}

%%%%%%%%%%%%%%%%%%%%%%%%%%%%%%%%%%%%%%%%%%%%%%%%%%%%%%%%%
\section{Introduction}

In statistical physics, it has been a central issue to clarify a relationship between the symmetry which breaks at the transition point and the nature of the phase transition such as the order of the phase transition and the universality class.
In order to consider such a relationship, the standard Potts model is often adopted\cite{Potts-1952,Wu-1982}.
In the standard Potts model, spins have $q$ values and its interaction is represented by Kronecker's delta.
Suppose we consider only two-body interaction, the Hamiltonian of the standard Potts model is given as
\begin{eqnarray}
 {\cal H}_{\rm s} = - \sum_{\langle i,j \rangle} J_{ij} \delta_{s_i,s_j},
  \,\,\,\,\,
  s_i = 1, \cdots, q,
\end{eqnarray}
where $\langle i,j \rangle$ denotes the pair of the interacted sites.
Note that the two-state Potts model is equivalent to the Ising model.
Then the $q$-state Potts model can be regarded as the standard generalization of the Ising model.
In the standard $q$-state ferromagnetic ($J_{ij} \ge 0$) Potts model on the two dimensional lattice, second-order and first-order phase transition with spontaneous breaking of $q$-fold symmetry occur when $q \le 4$ and otherwise, respectively.
The standard ferromagnetic Potts model has sometimes succeeded to explain the nature of phase transition in complicated theoretical models and real materials\cite{Wu-1982}.
For example, the phase transition of the absorption of $^3$He on graphite and the orbital order of transition metal oxides such as LaMnO$_3$ can be explained by three-state ferromagnetic Potts model\cite{Bretz-1977,Tejwani-1980,Ahmed-2009}.

Recently, there have been found novel examples that the nature of the phase transition is not consistent with that of the standard $q$-state Potts model\cite{Tamura-2008,Stoudenmire-2009,Okumura-2010}.
For instance, a first-order phase transition with spontaneous threefold symmetry breaking takes place in some two-dimensional frustrated systems with a number of competed short-range interactions.
This phase transition seems to be a counterexample of the concept of the standard ferromagnetic $q$-state Potts model, since the standard ferromagnetic three-state Potts model on two-dimensional lattice has a second-order phase transition as mentioned above.
Since mechanism of such a phase transition is not clear, from a viewpoint of statistical physics, it is important to understand why a first-order phase transition with threefold symmetry breaking occurs even on two-dimensional lattice.

Purpose of our study is to construct a simple model which can explain such a nature of phase transition by modifying the standard Potts model and to investigate a microscopic mechanism of changing the order of phase transition.
We found that the order of phase transition of three-state Potts model changes by adding invisible states\cite{Tamura-2010}.
In this paper, we study the detail of the phase transition of Potts model with invisible states and consider the effect of invisible states.
The organization of this paper is as follows.
In Section 2, we introduce the Potts model with invisible states and review some properties of this model.
In Section 3, we show the mean-field analysis and result obtained by Monte Carlo simulation.
In Section 4, we summarize our study and show future perspective.

\section{Model}

We consider the following Hamiltonian which is called the ($q$,$r$)-state Potts model\cite{Tamura-2010,Tanaka-2010a}:
\begin{eqnarray}
 \label{eq:original_Hamiltonian}
  {\cal H} = -J \sum_{\langle i,j \rangle}
  \delta_{s_i, s_j} 
  \sum_{\alpha = 1}^{q} \delta_{s_i, \alpha} \delta_{s_j, \alpha},
  \,\,\,\,\,\,\,
  s_i = 1, \cdots, q, q + 1, \cdots, q + r,
\end{eqnarray}
where $\langle i,j \rangle$ denotes the pair of the nearest neighbor sites on square lattice.
Note that spins can have discrete values from $1$ to $q+r$ in this model, while the spins in the standard $q$-state ferromagnetic Potts model can have a value from $1$ to $q$.
We call the states where $1 \le s_i \le q$ ``colored states'' and the states where $q+1 \le s_i \le q+r$ ``invisible states''\cite{Tamura-2010}.
In this paper, we only consider the case for ferromagnetic coupling $J > 0$.
If and only if $1 \le s_i = s_j \le q$, the interaction $J$ works and it is an energetically favorable state.

First we consider two spin system.
 \begin{figure}[t]
  \begin{center}
   \includegraphics[scale=0.85]{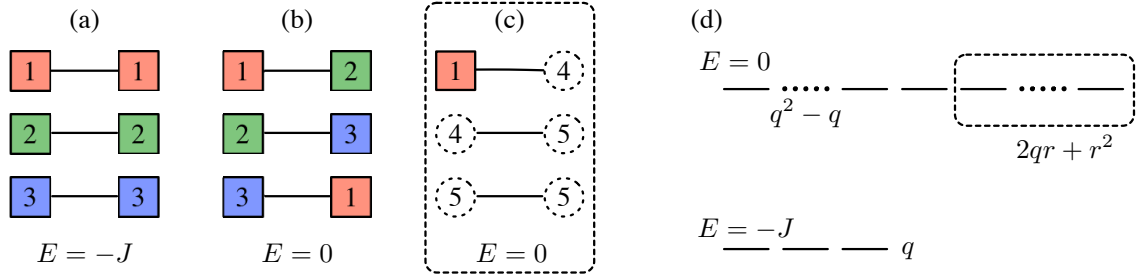}%
  \end{center}
  \caption{\label{fig:energy_twospins}
  (a)-(c): The ($3$,$2$)-state Potts model.
  The squares and dotted circles indicate colored state and invisible states, respectively.
  (a) Ground states of two spin system.
  (b) Excited states of two spin system.
  Both (a) and (b) also appear in the case of the standard ferromagnetic $q$-state Potts model.
  (c) Excited states of two spin system due to invisible states.
  (d) Energy level structure of two spin system for the ($q$,$r$)-state Potts model.
  The number of ground states and excited states are $q$ and $q^2-q+2qr+r^2$, respectively.
  The dotted boxes in (c) and (d) denote the contribution of invisible states.
  }
 \end{figure}
Figure \ref{fig:energy_twospins} (a) shows the ground states for the ($3$,$2$)-state Potts model.
Figure \ref{fig:energy_twospins} (b) and (c) denote excited states for the ($3$,$2$)-state Potts model.
States depicted in Fig.~\ref{fig:energy_twospins} (a) and (b) appear in the standard ferromagnetic $q$-state Potts model, whereas states depicted in Fig.~\ref{fig:energy_twospins} (c) appear only when the invisible states are introduced.
Figure \ref{fig:energy_twospins} (d) shows the energy level structure of two spin system for the ($q$,$r$)-state Potts model.
The number of ground states is $q$ and that of the excited states is $q^2 - q + 2qr + r^2$.
Contribution of the invisible states in excited states is $2qr + r^2$.
This energy level structure indicates that $q$-fold symmetry breaks at the transition point.

By comparing the partition function, we can obtain equivalent Hamiltonian:
\begin{eqnarray}
 \label{eq:equivalent_Hamiltonian}
  {\cal H}' = -J \sum_{\langle i,j \rangle} \delta_{\sigma_i, \sigma_j}
\sum_{\alpha = 1}^q \delta_{\sigma_i, \alpha} \delta_{\sigma_j, \alpha}
 - T \log r \sum_i \delta_{\sigma_i, 0},
 \,\,\,\,\,\,\,
 \sigma_i = 0,1,\cdots,q,
\end{eqnarray}
where we rename the label of invisible state from $q+1 \le s_i \le q+r$ to $\sigma_i = 0$.
The second term of Eq.~(\ref{eq:equivalent_Hamiltonian}) represents the chemical potential of invisible state.
The temperature-dependent chemical potential comes from entropy effect of invisible states.
Thus, the invisible state does not contribute the internal energy but affect the entropy.

\section{Result}

In the first half of this section, we consider the Bragg-Williams approximation.
In the last half of this section, we study a nature of phase transition by Monte Carlo simulation.

\subsection{Mean-field analysis}

First we consider the order of the phase transition of the Potts model with invisible states by Bragg-Williams approximation.
Let $x_\alpha$ be the fraction of $\alpha$-th state ($0 \le \alpha \le q$) with condition $\sum_{\alpha = 0}^{q} x_\alpha = 1$.
To analyzing the phase transition with spontaneous breaking of $q$-fold symmetry, we assume $x_\alpha$ as follows:
\begin{eqnarray}
 \label{eq:fraction_mf}
  x_0 = t,
  \,\,\,\,\,\,\,
  x_1 = \frac{1}{q}\left( 1 - t\right)\left[ 1 + \left( q - 1 \right) s \right],
  \,\,\,\,\,\,\,
  x_\alpha = \frac{1}{q}\left( 1 - t\right)\left( 1 - s\right) \,\,\,\,\, \left( 2 \le \alpha \le q\right),
\end{eqnarray}
where $0 \le s,t \le 1$.
Let $z$ be the number of the nearest neighbor sites.
The internal energy and the entropy are given by
\begin{eqnarray}
 \label{eq:energyentropy_bw}
 E^{\rm BW}\left( s, t \right) = -\frac{zJ}{2}\sum_{\alpha = 1}^q x_\alpha^2 - x_0 T\log r,
 \,\,\,\,\,\,\,
 S^{\rm BW}\left( s, t \right) = -\sum_{\alpha=0}^q x_\alpha \log x_\alpha,
\end{eqnarray}
respectively.
Then the free energy is given by
\begin{eqnarray}
 \label{eq:freeenergy_bw}
 F^{\rm BW}\left( s,t\right) &=& E_{\rm BW}\left( s,t\right) - T S_{\rm BW}\left( s,t\right) \\
 \nonumber
  &=& -\frac{zJ\left( 1- t\right)^2}{2q} \left[ \left( q-1\right)s^2 + 1\right] +t T\log \frac{t}{r} \\
  \label{eq:freeenergy_bw2}
  &+& (1-t)T 
  \left[
   \frac{1+(q-1)s}{q}\log\frac{1+(q-1)s}{1-s}
   + \log\frac{(1-t)(1-s)}{q}
  \right].
\end{eqnarray}
From this free energy, we can obtain the latent heat and transition temperature in the cases of $q=2,3$, and $4$.
 \begin{figure}[t]
  \begin{center}
   \includegraphics[scale=0.5]{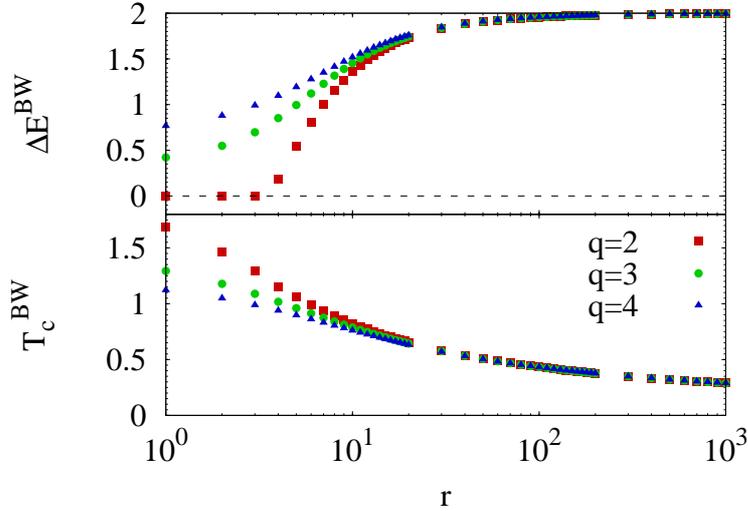}%
  \end{center}
 \caption{\label{fig:mf_tc_de}
  The transition temperature $T_{\rm c}^{\rm BW}$ and the latent heat $\Delta E^{\rm BW}$ as functions of the number of invisible states $r$ for $z=4$ obtained by the Bragg-Williams approximation.
}
 \end{figure}
Figure \ref{fig:mf_tc_de} shows $r$-dependency of the latent heat and transition temperature for $z=4$.
From Fig.~\ref{fig:mf_tc_de}, we find a second-order phase transition takes place only for ($2$,$1$), ($2$,$2$), and ($2$,$3$)-state Potts model.
Note that the transition temperature and the latent heat for $r=0$ ({\it i.e.} standard Potts model) are given as
\begin{eqnarray}
  T_{\rm c}^{\rm BW} \left( q=2, r= 0 \right) = \frac{zJ}{2},
   \,\,\,\,\,\,\,
 &&T_{\rm c}^{\rm BW} \left( q, r=0 \right) = 
  \frac{zJ}{2\log \left( q- 1\right)} \left( \frac{q-2}{q-1}\right) \,\,\, ({\rm for}\,\, q\ge 3),\\
 &&\Delta E^{\rm BW} \left( q, r=0 \right) = \frac{zJ\left( q- 2\right)^2}{2q\left( q- 1\right)} \,\,\, ({\rm for}\,\, q\ge 3),
\end{eqnarray}
respectively\cite{Kihara-1954}.
As the number of invisible states $r$ increases, the latent heat $\Delta E^{\rm BW}$ increases and the transition point $T_{\rm c}^{\rm BW}$ decreases.
From this fact, it can be considered that the invisible state drives a first-order phase transition.

\subsection{Monte Carlo simulation}

\begin{figure}[t]
 \begin{center}
  $
  \begin{array}{cccc}
  \includegraphics[scale=0.42]{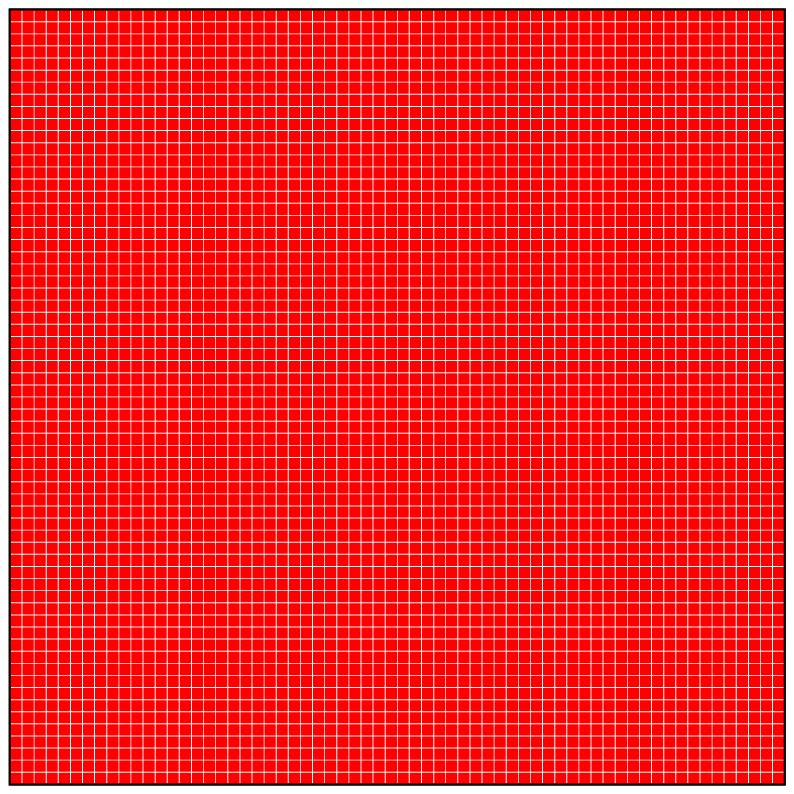}&
  \includegraphics[scale=0.42]{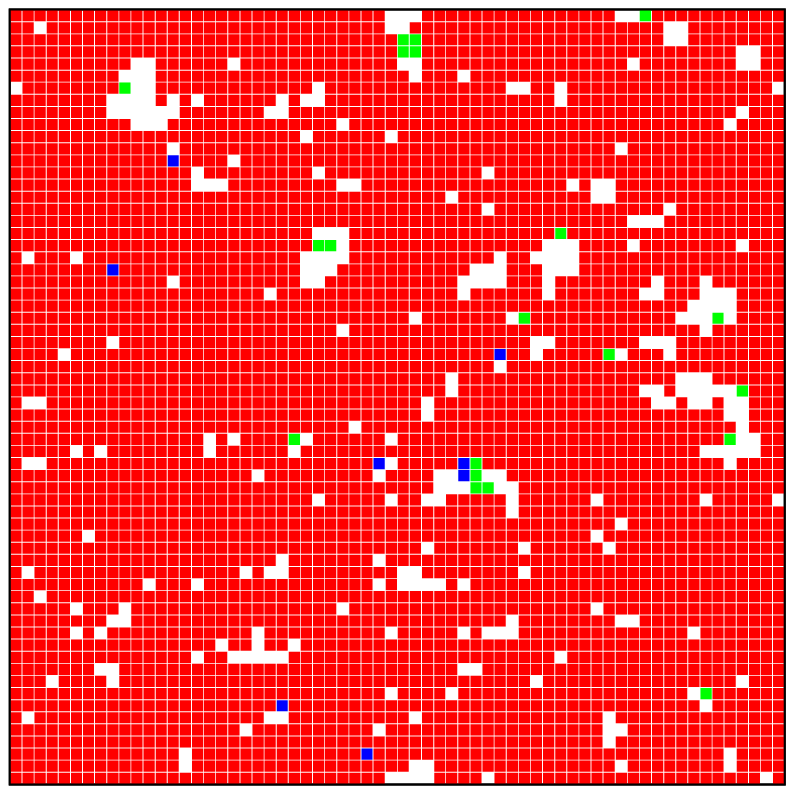}&
  \includegraphics[scale=0.42]{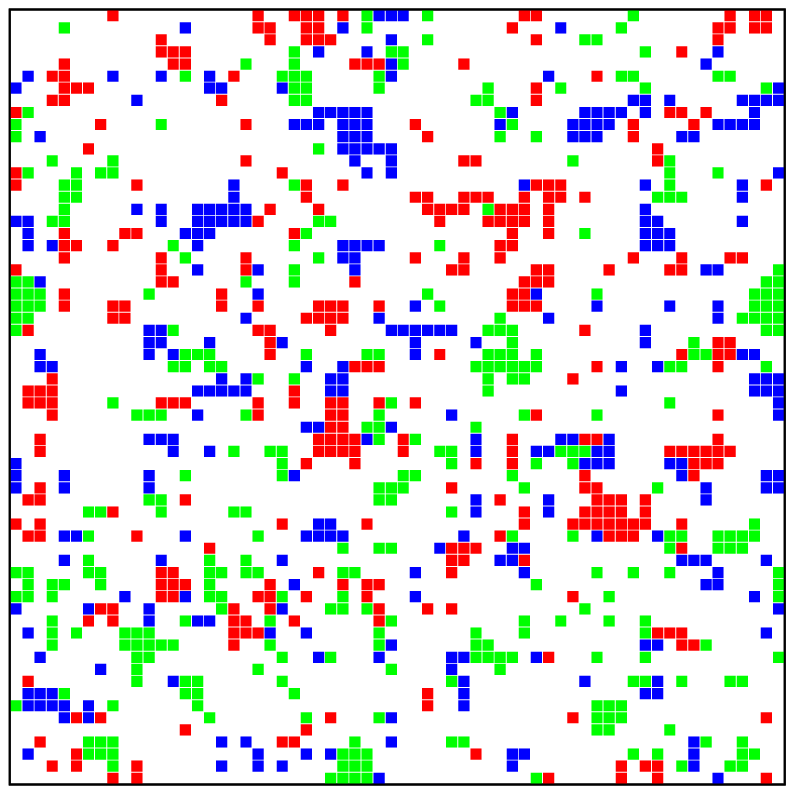}&
  \includegraphics[scale=0.42]{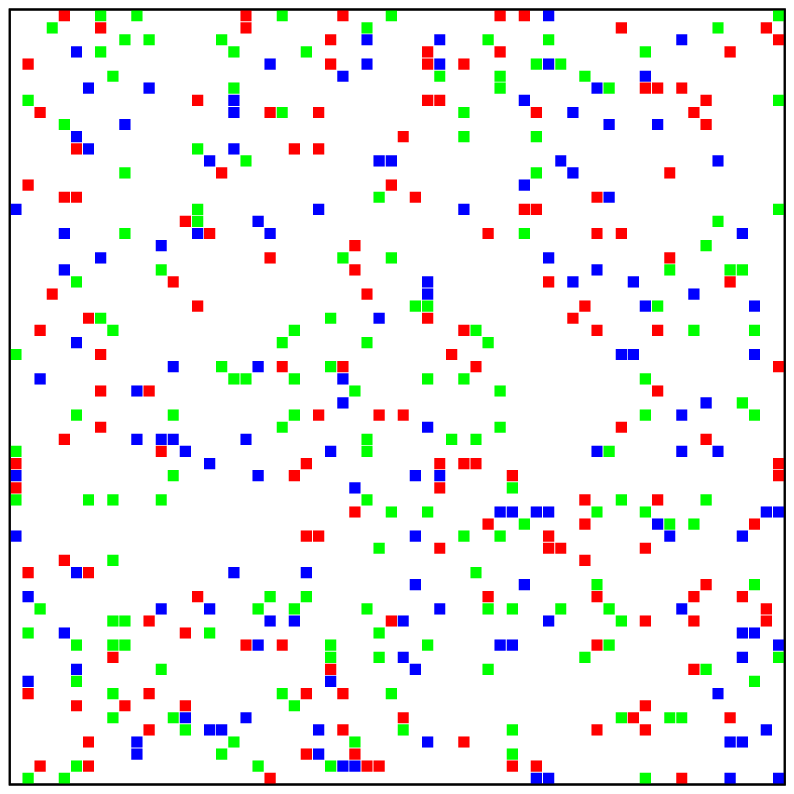}\\
   T=0 & T=0.58 & T=0.6 & T=1
  \end{array}
  $
  \caption{\label{fig:snapshot}
  Typical snapshots for the ($3$,$27$)-state ferromagnetic Potts model.
  As the temperature decreases, the density of invisible states decreases and the ordering domain grows.
  }
 \end{center}
\end{figure}

To investigate the effect of invisible states on two dimensional lattice, we study equilibrium properties of the ($q$,$r$)-state Potts model on square lattice by Monte Carlo simulation.
Figure \ref{fig:snapshot} shows typical snapshots of the ($3$,$27$)-state Potts model for several temperatures.
The color points and white points indicate the colored states and the invisible states, respectively.
The red, green, and blue points depict $\sigma_i = 1,2$, and $3$, respectively.
At zero temperature, all spins take the same colored state.
As the temperature increases, size of the ordered domains become small.
Above the transition temperature, small colored islands are floating in an invisible sea.
 \begin{figure}[b]
  \begin{center}
  $\begin{array}{ccc}
   \includegraphics[scale=0.5]{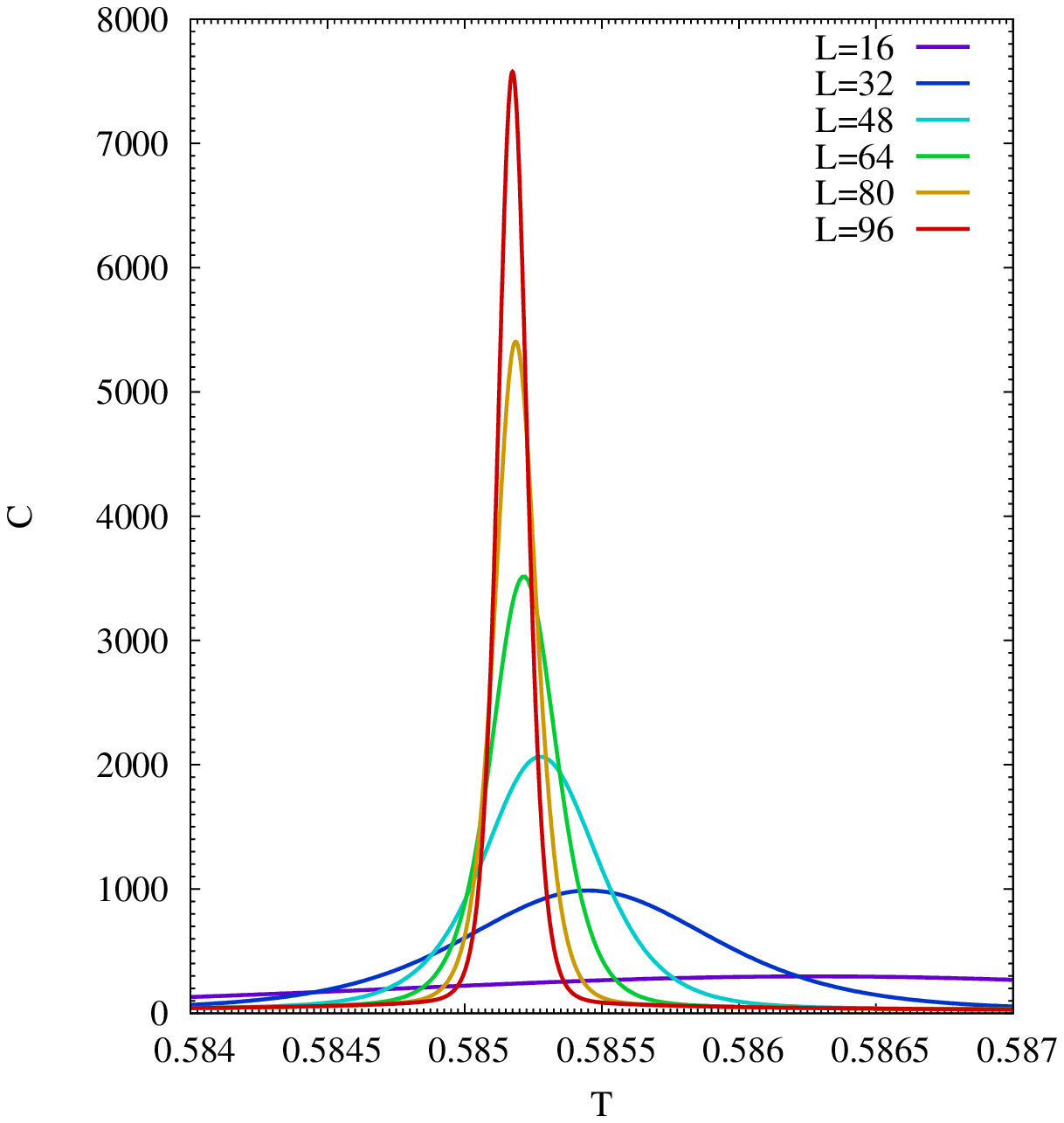}&%
    \hspace{5mm}&
   \includegraphics[scale=0.5]{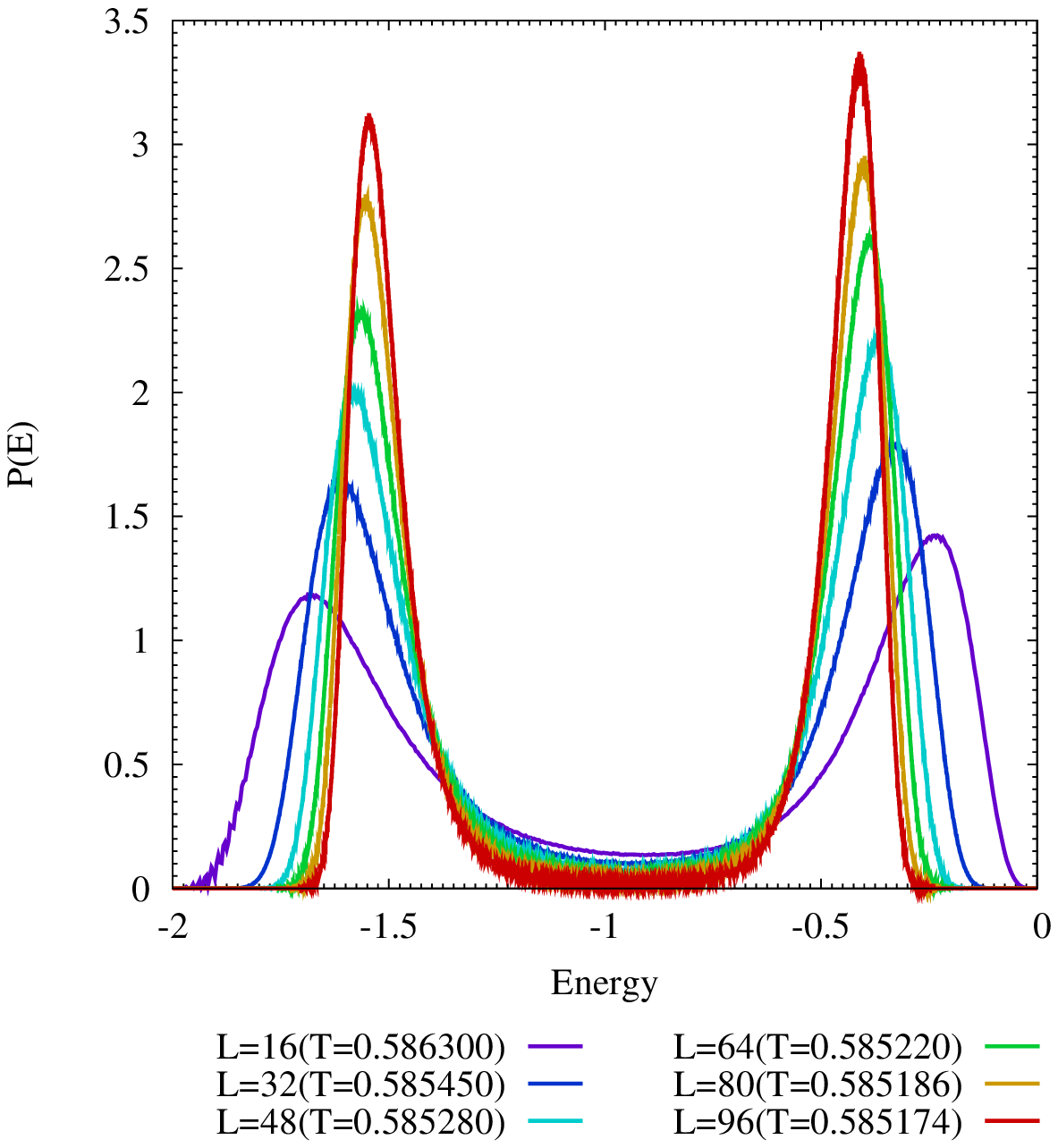}\\
     ({\rm a}) & \hspace{5mm} & ({\rm b})
   \end{array}
   $
   \end{center}
  \caption{\label{fig:eq_value}
  (a) Temperature dependency of the specific heat for the ($3$,$27$)-state Potts model.
  (b) Energy histogram for the ($3$,$27$)-state Potts model.
  }
 \end{figure}

Figure \ref{fig:eq_value} (a) shows the specific heat as a function of temperature.
In order to obtain the data with high accuracy, we adopt reweighting method.
The maximum value of specific heat increases almost linear dependence with system size.
To clarify whether the first-order phase transition occurs or not, next we calculate the energy distribution and implement finite size scaling.
Figure \ref{fig:eq_value} (b) shows the energy histogram at the temperature where the specific heat has the maximum value.
Energy histogram in Fig.~\ref{fig:eq_value} (b) is a bimodal distribution which is a characteristic behavior of the first-order phase transition.
As the system size increases, the peaks become high.

\begin{figure}[h]
  \begin{center}
   $\begin{array}{ccc}
   \includegraphics[scale=0.6]{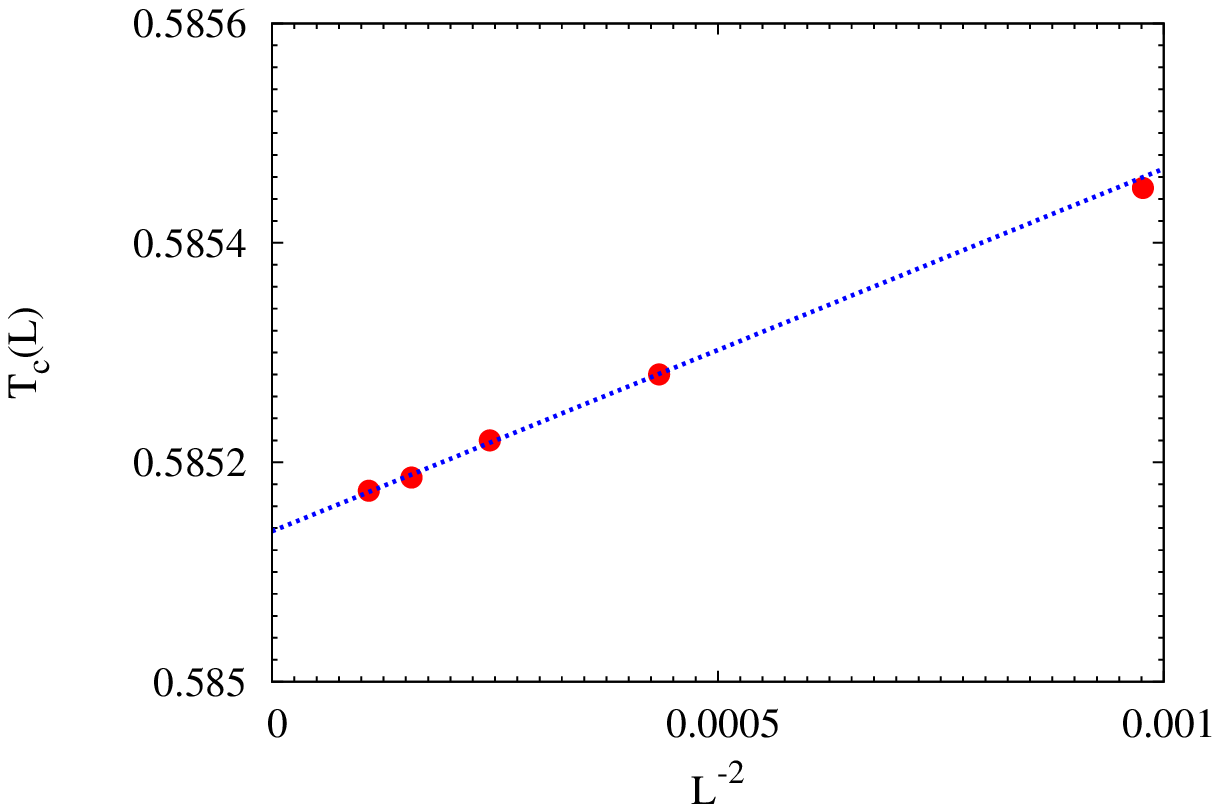}&%
    \hspace{5mm}&
   \includegraphics[scale=0.6]{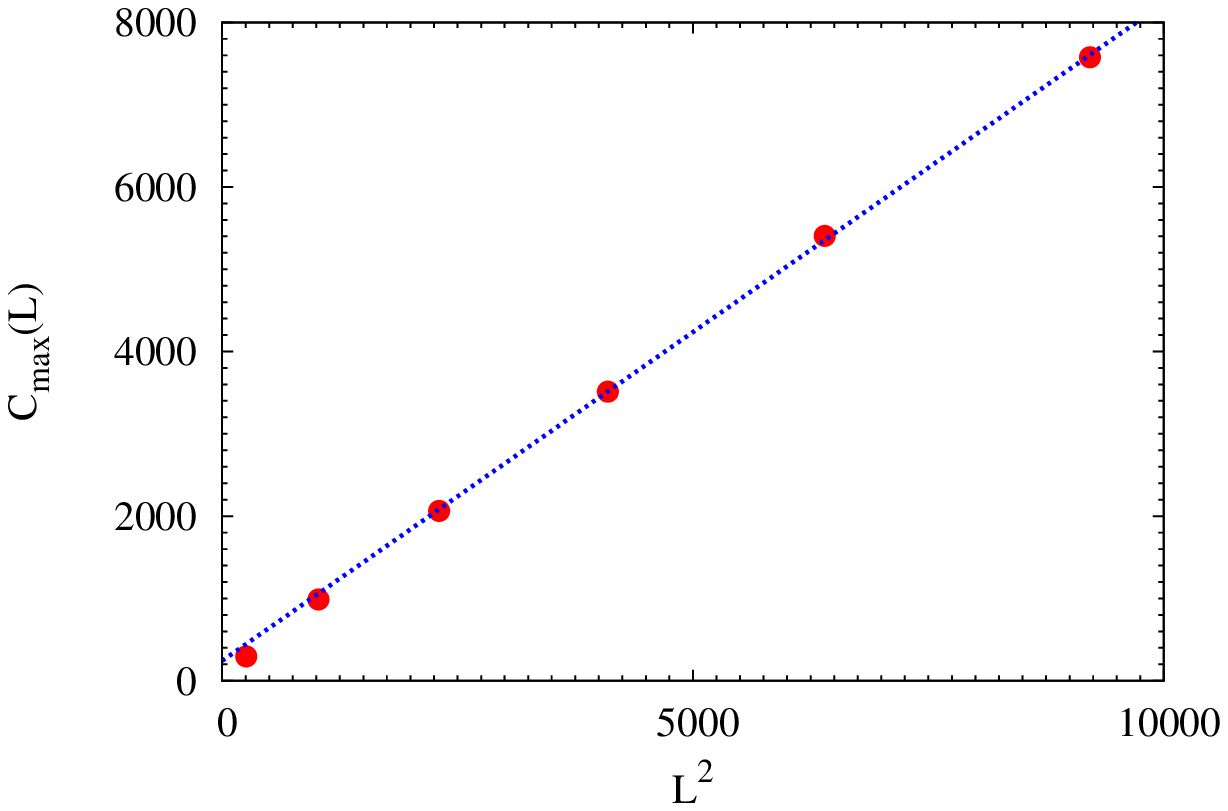}\\%
   ({\rm a}) & \hspace{5mm} & ({\rm b})
   \end{array}$
  \end{center}
  \caption{\label{fig:fitting}
(a) $T_{\rm c}\left( L \right)$ as a function of $L^{-2}$. Intercept of the fitting curve indicates the estimated transition temperature for the thermodynamic limit.
(b) $C_{\rm max}\left( L \right)$ as a function of $L^2$. 
}
 \end{figure}

Next we implement the finite size scaling which is established by Challa {\it et al.}\cite{Challa-1986}. 
The finite size scaling for a first-order phase transition in $d$-dimensional system is treated as:
\begin{eqnarray}
 T_{\rm c} \left( L \right) = aL^{-d} + T_{\rm c}\left( \infty\right),
  \,\,\,\,\,\,\,
 C_{\rm max} \left( L \right) \propto \frac{\left( \Delta E\right)^2 L^d}{4 T_{\rm c}^2 \left( \infty\right)},
\end{eqnarray}
where $T_{\rm c}\left( L \right)$ denotes the temperature where the specific heat takes the maximum value $C_{\rm max} \left( L \right)$ in finite size $L^d$ system.
The values $T_{\rm c}\left( \infty \right)$ and $\Delta E$ denote the transition temperature and the latent heat for the thermodynamic limit, respectively.
We fit $T_{\rm c} \left( L \right)$ as a function of $L^{-2}$ as shown in Fig.~\ref{fig:fitting}(a).
From the intercept of fitting curve, we obtain $T_{\rm c}\left( \infty \right) = 0.58513(1)$.
Next we fit $C_{\rm max} \left( L \right)$ as a function of $L^2$.
We estimate the latent heat $\Delta E = 1.05(2)$ from the gradient of the fitting curve and the obtained transition temperature $T_{\rm c}\left( \infty\right)$ as shown in Fig.~\ref{fig:fitting}(b).

By the same analysis, we calculate the transition temperature and the latent heat for several parameter set of ($q$,$r$) (Table \ref{table:tc_lh}).
As the number of invisible states increases, the transition temperature decreases and latent heat increases.
From Table~\ref{table:tc_lh}, we conclude the invisible states stimulate a first-order phase transition, which is consistent with the result obtained by Bragg-Williams approximation.
\begin{table}
\begin{center}
 \caption{\label{table:tc_lh}
 Transition temperature and latent heat for several parameter set of ($q$,$r$)-state Potts model.
 For $r=0$ case, $T_{\rm c}$ and $\Delta E$ were obtained exactly\cite{Wu-1982}.}
 \begin{tabular}{c|ccc}
  \hline\hline
  ($q$,$r$) & $T_{\rm c}$ & $\Delta E$ & Symmetry \\
  \hline
  ($2$,$0$) & $1.13459$ & $0$ & 2-fold \\
  ($2$,$30$) & $0.57837(1)$ & $1.02(2)$ & 2-fold \\
  ($2$,$32$) & $0.56857(1)$ & $1.23(2)$ & 2-fold \\
  ($3$,$0$) & $0.994973$ & $0$ & 3-fold \\
  ($3$,$25$) & $0.59630(1)$ & $0.81(2)$ & 3-fold \\
  ($3$,$27$) & $0.58513(1)$ & $1.05(2)$ & 3-fold \\
  ($4$,$0$) & $0.910239$ & $0$ & 4-fold \\
   ($4$,$20$) & $0.61683(1)$ & $0.68(2)$ & 4-fold \\
  ($4$,$22$) & $0.60396(1)$ & $0.87(2)$ & 4-fold \\
  \hline\hline
 \end{tabular} 
\end{center}
\end{table}

\section{Conclusion and Future Perspective}

In this paper, we study the nature of the phase transition of the Potts model with invisible states.
The Potts model with invisible states is regarded as the straightforward extension of the standard ferromagnetic Potts model, since this model is constructed by just adding the invisible states.
The invisible state does not affect the internal energy but contributes to the entropy.
It does not change the symmetry which breaks at the transition temperature.
From the result obtained by Bragg-Williams approximation and Monte Carlo simulation, the invisible state drives a first-order phase transition, since the latent heat increases as the number of the invisible states increases.
We believe that this model is a fundamental model and it enables us to understand why the first-order phase transition with spontaneous $2$,$3$, and $4$-fold symmetry breaking can occur even on two-dimensional lattice.

Recently, the Potts model has been adopted for analysis of problems in information science and technology such as clustering problem\cite{Kurihara-2009,Sato-2009}.
It has been suggested that the effects of thermal and quantum fluctuation are efficient for optimization problems.
We expect that the fluctuation which comes from the invisible states is efficient for such problems.

%%%%%%%%%%%%%%%%%%%%%%%%%%%%%%%%%%%%%%%%%%%
%\appendix
%%%%%%%%%%%%%%%%%%%%%%%%%%%%%%%%%%%%%%%%%%%%%%
%\section{Derivation of overlap function:  A dynamical approach}
%\label{sec:AppA}
%%%%%%%%%%%%%%%%%%%%%%%%%%%%%%%%%%%%%%%%% 
%In this appendix, we show the derivation 
%%%%%%%%%%%%%%%%%%%
%%%%%%%%%%%%%%%%%%%%%%%%%%%%%%%%%%%%%%%%%%%%%%%%%%%%%%%%%%%%%%%%%%%%%%
\ack
%%%%%%%%%%%%%%%%%%%%%%%%%%%%%%%%%%%%%%%%%%%%%%%%%%%%%%%%%%%%%%%%%%%%%% 
The authors are grateful to Jie Lou, Yoshiki Matsuda, Seiji Miyashita, Takashi Mori, Yohsuke Murase, Masayuki Ohzeki, and Eric Vincent for their valuable comments. 
S.~T. is partly supported by Grant-in-Aid for Young Scientists Start-up (21840021) from the JSPS and the ``Open Research Center'' Project for Private Universities: matching fund subsidy from MEXT.
R.T. is partly supported by Global COE Program ``the Physical Sciences Frontier'', MEXT, Japan. 
The present work is financially supported by MEXT Grant-in-Aid for Scientific Research (B) (22340111), and for Scientific Research on Priority Areas ``Novel States of Matter Induced by Frustration'' (19052004), and by Next Generation Supercomputing Project, Nanoscience Program, MEXT, Japan. 
The computation in the present work was performed on computers at the Supercomputer Center, 
Institute for Solid State Physics, University of Tokyo.
%%%%%%%%%%%%%%%%%%%%%%%%%%%%%%%%%%%%%%%%%%%%

\end{document}